\UseRawInputEncoding
\documentclass[twocolumn,english,prb]{revtex4-2}
\usepackage[T1]{fontenc}
\usepackage{geometry}
\usepackage{verbatim}
\usepackage{amsmath}
\usepackage{amssymb}
\usepackage{graphicx}
\usepackage{esint}

\makeatletter
\@ifundefined{date}{}{\date{}}
\usepackage[caption=false]{subfig}

\@ifundefined{showcaptionsetup}{}{%
 \PassOptionsToPackage{caption=false}{subfig}}
\usepackage{subfig}
\makeatother

\usepackage{babel}
\begin{document}
\title{Proximity-induced equilibrium supercurrent and perfect superconducting
diode effect due to band asymmetry}
\author{Pavan Hosur}
\affiliation{Department of Physics and Texas Center for Superconductivity, University
of Houston, Houston, Texas 77004}
\author{Daniel Palacios}
\affiliation{Jan and Dan Duncan Neurological Research Institute at Texas Children's Hospital, Houston, TX, 77030}
\affiliation{Graduate Program of Quantitative and Computational Biosciences, Baylor College of Medicine, Houston, TX, 77030}
\begin{abstract}
We theoretically investigate the consequences of proximity-induced
conventional superconductivity in metals that break time-reversal
and inversion symmetries through their energy dispersion. We discover
behaviors impossible in an isolated superconductor such as an equilibrium
supercurrent that apparently violates a no-go theorem and, at suitable
topological defects, non-conservation of electric charge reminiscent
of the chiral anomaly. The equilibrium supercurrent is expected to
be trainable by a helical electromagnetic field in the normal state.
Remarkably, if the band asymmetry exceeds the critical current of
the parent superconductor in appropriate units, we predict a perfect
superconducting diode effect with diode coefficient unity. We propose
toroidal metals such as UNi$_{4}$B and metals with directional scalar
spin chiral order as potential platforms.
\end{abstract}
\maketitle

\section{Introduction}
Nonreciprocal phenomena in superconductors (SCs) have a long history. Among
diode-like systems, early examples included amplification of the luminescence
of light-emitting diodes when the diode was attached to a SC \citep{asano2003theoretical,Hayashi_2008}. More recently, the
asymmetry in the current-voltage characteristics of non-centrosymmetric
metals under a magnetic field was seen to be enhanced if the metal
turned superconducting \citep{Wakatsuki2017,Hoshino2018,Wakatsuki2018}.
Recent theoretical and experimental breakthroughs in the theory and
realization of superconducting and Josephson diodes \citep{Wakatsuki2017,Hoshino2018,Wakatsuki2018,Yasuda:2019wb,Ando:2020td,Itahashi2020,Diez-Merida2021,Miyasaka_2021,Baumgartner:2022wr,Baumgartner_2022,Bauriedl:2022we,Daido2022,Daido2022a,Golod:2022ta,Karabassov2022,He_2022,Narita:2022tb,Pal:2022tm,wang2022symmetry,Wu:2022wq,Yuan2022,Zhai2022,Zhang2022,Satchell2023},
which carry immense technological potential by avoiding the enormous
heating losses of semiconductor diodes, have driven fervent activity
in the field. These diodes are characterized by unequal critical supercurrents
in opposite directions, resulting in Ohmic and dissipationless transport,
respectively, for current magnitudes between the two critical currents.
Such diode effects are intimately connected to the exotic Fulde-Ferrell
superconductivity, defined by finite momentum Cooper pairs in the
ground state \citep{Ando:2020td,Lin2021,Bauriedl:2022we,Daido2022,He_2022,Yuan2022}.
Another exotic non-reciprocal phenomenon entails the existence of
spontaneous supercurrents in a preferred direction through Josephson
junctions \citep{Amin2001,Braude2007,Ashby2009,Heim_2013,Goldbin2015,Bobkova2016,Szombati:2016aa,Alidoust2017,Alidoust2018,Alidoust2018a,Alidoust2018b,Malshukov2018,Assouline:2019aa,Shukrinov2019,Alidoust2020,Alidoust2020a,Kulikov2020,Mazanik2020,Sinha2020,Alidoust2021,Liu_2021,Halterman2022,Monroe2022,Meng2022,Satchell2023,Xie2023}
and SCs with spin-orbit coupling in proximity to magnetism \citep{Pershoguba2015,Malshukov2016,Malshukov2017,Mironov2017,Robinson2019,Samokhvalov:2021vh}.
While details vary, all the above approaches rely crucially on one
principle: broken $\mathcal{T}$ and $\mathcal{I}$ symmetries. Violation
of these symmetries results in other peculiar phenomena, such as unusual
vortex dynamics in non-centrosymmetric SCs \citep{Miclea2009,Miclea2010,bauer2012non,Cameron2019}.

In this work, we revisit the problem of non-reciprocity in superconducting
systems and explore it in a minimal scenario. Specifically, we consider\emph{
}metals with an asymmetric dispersion $\varepsilon_{\boldsymbol{k}}\neq\varepsilon_{-\boldsymbol{k}}$
and no Berry phases, proximity-couple them to a conventional, $s$-wave
SC and focus on a uniform system without any Josephson
junctions. We dub metals with $\varepsilon_{\boldsymbol{k}}\neq\varepsilon_{-\boldsymbol{k}}$
\emph{band asymmetric metals} (BAMs) and refer to BAMs that acquire
conventional superconductivity as band asymmetric superconductors
(BASCs). Since $\boldsymbol{k}$ is inequivalent to $-\boldsymbol{k}$
in BAM, intrinsic pairing tendencies in them, if any, are expected
to be towards exotic Fulde-Ferrell superconductivity built from finite
momentum Cooper pairs \citep{FuldeFerrell}. On the other hand, band
asymmetry eliminates a Cooper instability at weak interactions, so
a more practical route to superconductivity in BAMs may be extrinsic.
We show that even this minimal setup leads to strange behaviors impossible
in isolated SCs, namely, (i) an apparent violation of a basic no-go theorem due to an equilibrium current density (Sec. \ref{sec:eqbm-sc}), (ii) a perfect superconducting
diode effect (SDE) without fine-tuning (Sec. \ref{sec:perfect}), and (iii) topological defects
that violate charge conservation (Sec. \ref{sec:defects}). We conclude by mentioning suitable experimental platforms (Sec. \ref{sec:platforms}).

\section{Equilibrium supercurrent} \label{sec:eqbm-sc}

We first derive the equilibrium current $I^{\text{eq}}$
in a one-dimensional (1D) BAM. Generalization to higher dimensions
is straightforward. While spontaneous supercurrents have been studied
\citep{Amin2001,Braude2007,Ashby2009,Heim_2013,Goldbin2015,Pershoguba2015,Bobkova2016,Malshukov2016,Szombati:2016aa,Alidoust2017,Malshukov2017,Mironov2017,Alidoust2018,Alidoust2018a,Alidoust2018b,Malshukov2018,Assouline:2019aa,Robinson2019,Shukrinov2019,Alidoust2020,Alidoust2020a,Kulikov2020,Mazanik2020,Sinha2020,Alidoust2021,Liu_2021,Samokhvalov:2021vh,Halterman2022,Monroe2022,Meng2022,Satchell2023,Xie2023},
their significance with respect to basic quantum mechanics has not
been appreciated, which we do here. In particular, we show how $I^\text{eq}$ naively violates a theorem by Bloch that forbids current densities in the thermodynamic limit in arbitrary systems of interacting fermions \citep{Bohm1949,Ohashi1996,Yamamoto2015,Tada:2016aa,Bachmann:2020aa}, and then resolve the paradox.

\begin{figure}
\includegraphics[width=0.9\columnwidth]{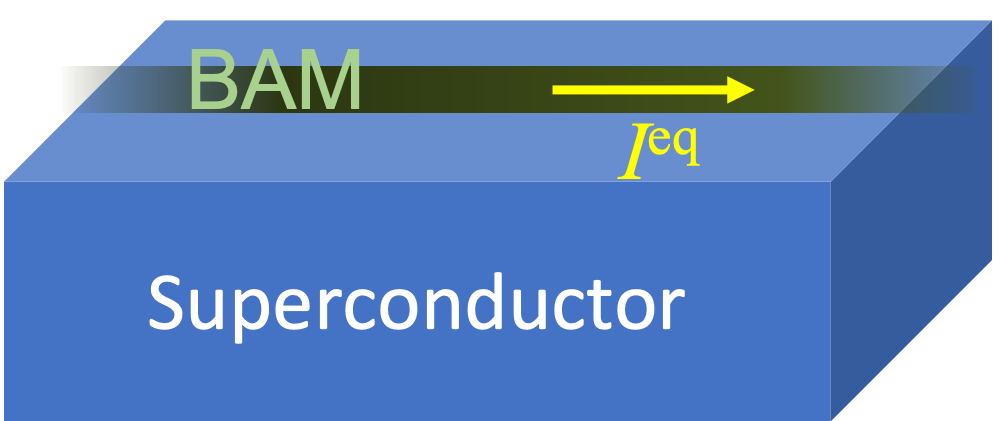}

\caption{Depositing a BAM wire on a conventional SC will generate an equilibrium
current $I^{\text{eq}}$, a SDE in general, and a perfect SDE with
unit diode coefficient if the band asymmetry exceeds a threshold determined
by the critical Cooper pair momentum of the parent SC. (See text for
details) \label{fig:BASC wire}}
\end{figure}

We assume a single band with degeneracy $g$; for spin-degenerate
bands, $g=2$. The Bloch Hamiltonian for such a BAM is 
\begin{equation}
H_{\text{BAM}}=\intop_{k}\sum_{n=1}^{g}c_{kn}^{\dagger}c_{kn}\varepsilon_{k}
\end{equation}
where $\intop_{k}\equiv\int\frac{dk}{2\pi}$. Let us deposit the BAM
wire on a conventional, $s$-wave SC with zero Cooper pair momentum,
as sketched in Fig~\ref{fig:BASC wire}. The BAM will develop conventional
superconductivity too via the proximity effect. The Bogoliubov-deGennes
Hamiltonian in the basis $\Psi_{k}=\left(c_{k}^{T},\mathcal{T}c_{k}^{\dagger}\mathcal{T}^{-1}\right)^{T}$
is $H_{\text{BdG}}=\frac{1}{2}\sum_{k}\Psi_{k}^{\dagger}\left(H_{k}^{\Delta}\otimes\mathbb{I}_{g}\right)\Psi_{k}$
where

\begin{align}
H_{k}^{\Delta} & =\left(\begin{array}{cc}
\varepsilon_{k} & \Delta_{0}^{*}\\
\Delta_{0} & -\varepsilon_{-k}
\end{array}\right)\label{eq:H-Delta}
\end{align}
and $\mathbb{I}_{g}$ is a $g\times g$ identity matrix. $I^{\text{eq}}$
is given by
\begin{equation}
I^{\text{eq}}=\intop_{k}\text{Tr}\left\{ j_{k}\left[f\left(H_{k}^{\Delta}\right)-f\left(H_{k}^{0}\right)\right]\right\} \label{eq:current}
\end{equation}
where $j_{k}=\frac{e}{2}\left(\begin{array}{cc}
v_{k} & 0\\
0 & -v_{-k}
\end{array}\right)\otimes\mathbb{I}_{g}$ is the current operator, $f(X)=\left[e^{X/T}+1\right]^{-1}$ and
we have set $\hbar=k_{B}=1$. We have explicitly subtracted a spurious
current due to Hilbert space doubling that captures the current carried
by the filled bands when $\Delta=0$. This current vanishes in general
lattice models and in continuum models with a symmetric dispersion.
However, in an asymmetric continuum, it is non-zero, regularization-dependent,
and can even diverge. For weak pairing, we find (Appendix \ref{app:Iq})
\begin{equation}
I^{\text{eq}}\approx ge|\Delta_{0}|^{2}\intop_{k}\frac{v_{-k}}{\left(\varepsilon_{k}+\varepsilon_{-k}\right)^{2}}\tanh\left[\frac{\varepsilon_{k}}{2T}\right]\label{eq:current answer}
\end{equation}
to leading order in $\Delta_{0}$. $I^{\text{eq}}$ is generically
non-zero as long as $\varepsilon_{k}\neq\varepsilon_{-k}$. To gain
more insight into this result, suppose the BAM has Fermi momenta $K_{i}$
and Fermi velocities $v_{i}$. Linearizing the dispersion as $\varepsilon_{K_{i}+p}\approx v_{i}p$,
$\varepsilon_{-K_{i}+p}\approx\varepsilon_{-K_{i}}$ and assuming
$\left|\varepsilon_{-K_{i}}\right|\gg\Lambda$ where $\Lambda$ is
an energy cutoff gives
\begin{equation}
I^{\text{eq}}\approx\frac{ge|\Delta_{0}|^{2}\Lambda^{2}}{2\pi}\sum_{i}\frac{v_{-K_{i}}}{\left|v_{i}\right|\varepsilon_{-K_{i}}^{3}}
\end{equation}
for $T\to0$. If we assume $\sum_{i}\frac{\Lambda^{2}v_{-K_{i}}}{\left|v_{i}\right|\varepsilon_{-K_{i}}^{3}}\sim10^{-9}/$eV,
which amounts to a 1 part-per-million band asymmetry if $\Lambda$
and $\varepsilon_{-K_{i}}$ are each $O($meV$)$ and all Fermi velocities
are of the same order, then $\Delta_{0}\sim1$K gives a large $I^{\text{eq}}\sim10$mA
which should be detectable via the magnetic fields it creates. 

The above current seems to contradict a seminal theorem by Bloch,
which states that the ground or equilibrium state of a generic, interacting
fermionic system cannot carry a current density \citep{Bohm1949,Ohashi1996,Yamamoto2015,Tada:2016aa,Bachmann:2020aa}.
In particular, a recent refinement of the theorem showed that the
current density along $x$ is bounded as $\left|\left\langle J_{x}\right\rangle \right|<O\left(L_{x}^{-1}\right)$,
where $L_{x}$ is the linear dimension in the $x$ direction \citep{Watanabe:2019aa}.
Historically, Bloch's theorem helped prove that persistent currents
in isolated superconducting and metallic \citep{Bleszynski-Jayich272,Bluhm2009}
rings necessarily occur in excited states and are stabilized by the
quantization of magnetic flux piercing the ring. Thus, the persistent
currents there have a long lifetime that is limited only by the probability
of spontaneous or stimulated emission that relaxes them to the ground
state. In contrast, BASCs clearly carry a ground state current with
a truly infinite lifetime, apparently evading Bloch's theorem. The
spontaneous supercurrents described in Refs. \citep{Pershoguba2015,Malshukov2016,Malshukov2017,Mironov2017,Robinson2019,Samokhvalov:2021vh}
are special cases of $I^{\text{eq}}$. However, $I^{\text{eq}}$ differs
fundamentally from spontaneous currents in $\mathcal{T}$ and $\mathcal{I}$
breaking Josephson junctions that crucially rely on the presence of
a junction and decay exponentially with junction thickness \citep{Amin2001,Braude2007,Ashby2009,Heim_2013,Goldbin2015,Bobkova2016,Szombati:2016aa,Alidoust2017,Alidoust2018,Alidoust2018a,Alidoust2018b,Malshukov2018,Assouline:2019aa,Shukrinov2019,Alidoust2020,Alidoust2020a,Kulikov2020,Mazanik2020,Sinha2020,Alidoust2021,Liu_2021,Halterman2022,Monroe2022,Meng2022,Satchell2023,Xie2023}
while $I^{\text{eq}}$ is independent of the length of the BAM wire.

The resolution to the paradox lies in the observation that Bloch's
theorem explicitly assumes charge conservation whereas the BASC can
freely exchange pairs of electrons with the parent SC. Viewed differently,
the BAM-plus-SC system conserves charge, obeys Bloch's theorem and
indeed has a vanishing current density in the thermodynamic limit.
However, the BASC alone can host a non-zero current density, which
physically corresponds to a surface current for the combined system
and is not suppressed by Bloch's theorem. Yet another interpretation
of the result is that the superconducting instability of an isolated
BAM is towards a finite momentum state. In other words, a $q\neq0$
pairing state minimizes the Ginzburg-Landau free energy, or equivalently,
solves the superconducting mean-field equations self-consistently,
of the isolated BAM wire with suitable interactions at low temperatures.
Then, the induced $q=0$ pairing state can be viewed as an excited
state of an isolated superconducting BAM and is therefore not restricted
by Bloch's theorem.

It is instructive to contrast the above current with topological boundary
phenomena. In particular, topological condensed matter physics is
rife with phenomena that are forbidden in isolated systems, but occur
robustly on the boundaries of topological phases. There, the violation
of the relevant no-go theorems on one boundary is cured by the opposite
boundary. From this perspective, the above current is a \emph{non-topological}
phenomenon that is forbidden in an isolated SC, but occurs robustly
on the surface of a conventional SC. Here, the apparent violation
of the relevant no-go theorem is rectified by the parent SC that acts
as an infinite reservoir of Cooper pairs.

\section{Perfect SDE}\label{sec:perfect}

We now argue that the above system realizes a
perfect superconducting diode for large enough band asymmetry with diode coefficient at its theoretical maximum, $\eta=1$, while small band asymmetry still results in a non-zero $\eta$. Unlike, for instance, Ref. \cite{Yuan2022}, where perfect diode behavior requires fine-tuning to a tricritical point, the perfect behavior here appears immediately once the band asymmetry exceeds a threshold. This remarkable behavior directly aligns with the central pursuit of the field of achieving a large $\eta$. While experimental non-idealities such as contact resistance will undoubtedly reduce $\eta$ in our proposal, the fact that the ideal scenario predicts $\eta=1$ without fine-tuning is exciting. In comparison, the largest $\eta$ experimentally achieved so far is $\eta\approx0.35$ in a heterostructure of $\beta$-Sn superconducting nanowires embedded in $\alpha$-Sn Dirac semimetal \cite{ishihara2023giant}.

\begin{figure}
\subfloat[]{\includegraphics[width=0.48\columnwidth]{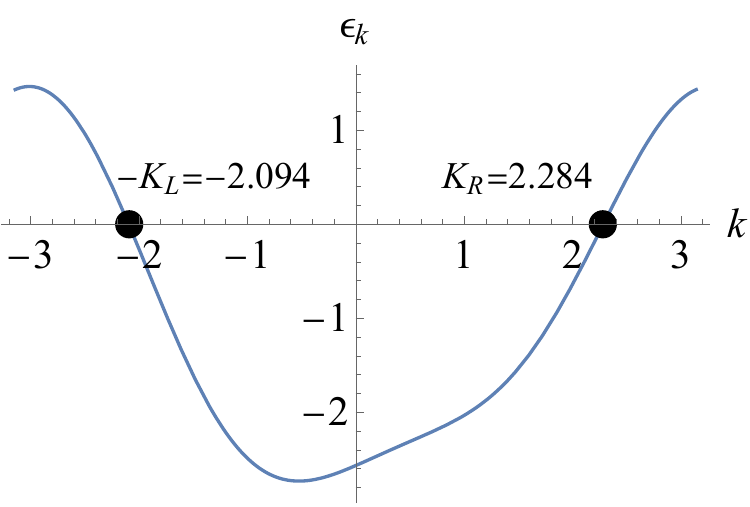}

}\subfloat[]{\includegraphics[width=0.48\columnwidth]{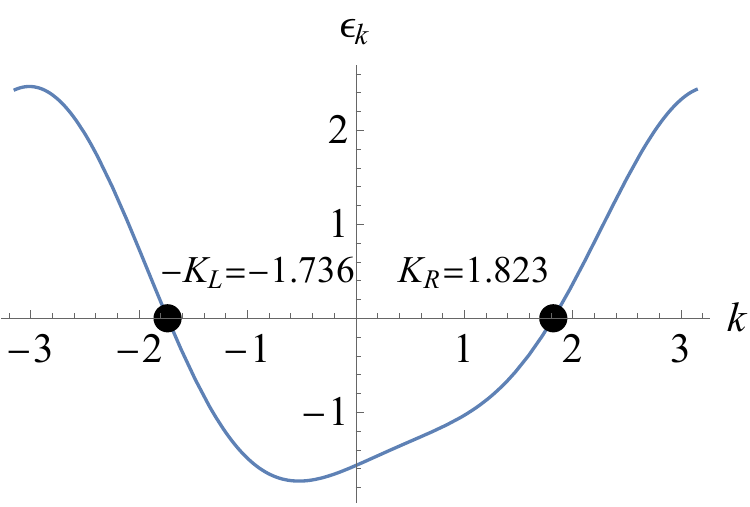}

}

\subfloat[]{\includegraphics[width=0.48\columnwidth]{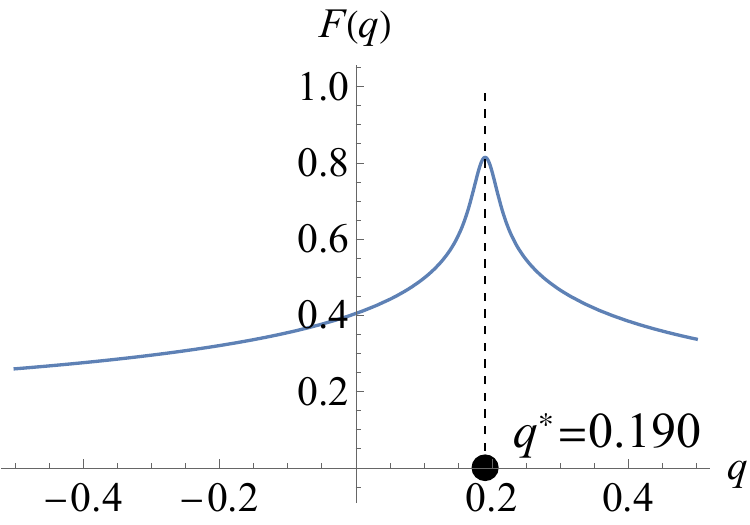}

}\subfloat[]{\includegraphics[width=0.48\columnwidth]{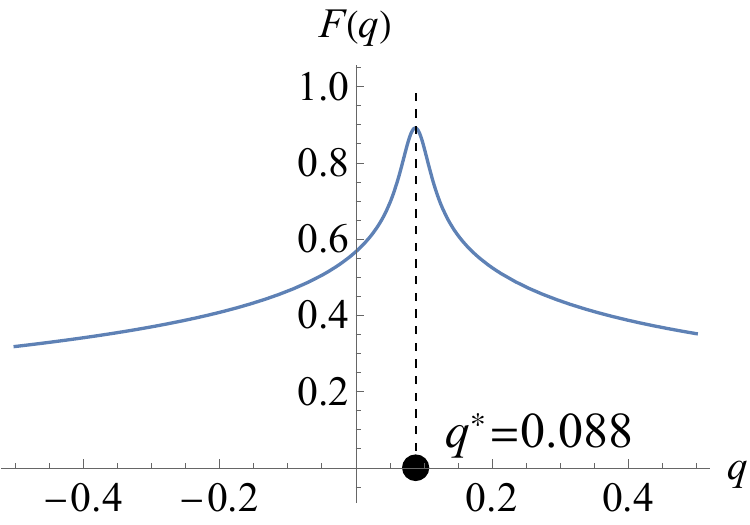}

}

\subfloat[]{\includegraphics[width=0.48\columnwidth]{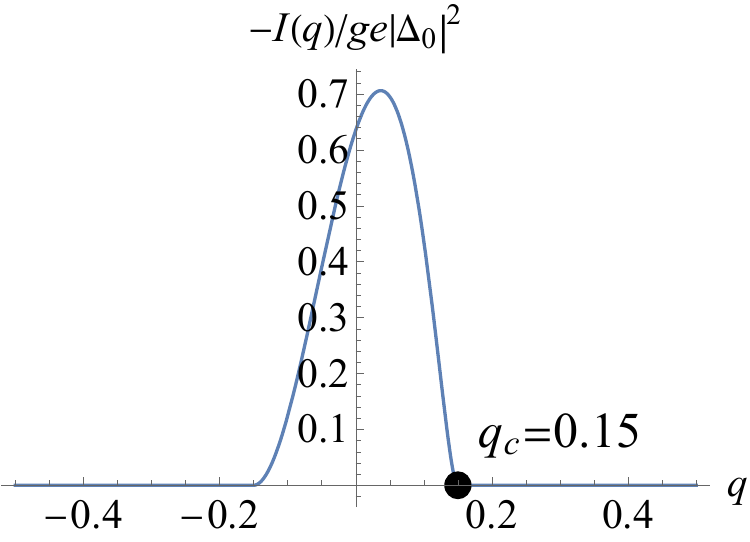}

}\subfloat[]{\includegraphics[width=0.48\columnwidth]{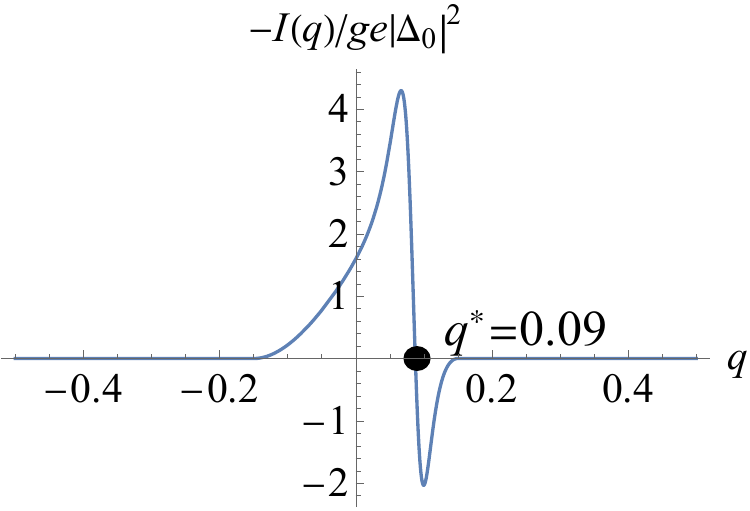}

}

\caption{An asymmetric dispersion (a,b), the corresponding $F(q)$ (c,d) and
the resulting $I(q)$ following Eq. (\ref{eq:F(q)}). We use $\varepsilon_{k}=-2\cos k+0.5\sin(2k+\pi/3)-\mu$
with $\mu=1$ for (a,c,e) and $\mu=0$ for (b,d,f), set $T=0.01$
and phenomenologically choose $\Delta_{q}=\Delta_{0}(1-q^{2}/q_{c}^{2})$
with $q_{c}=0.15$. Note that $q^{*}=K_{R}-K_{L}$ to very good accuracy
in both columns. When $|q^{*}|>q_{c}$ (left column), $I(q)$ is always
positive and vanishes at $q=q_{c}$ resulting in a perfect SDE. If
$|q^{*}|<q_{c}$, $I(q)<0$ for $|q|\in(|q^{*}|,q_{c})$ and the SDE
is imperfect.\label{fig:SDE}}
\end{figure}

The proof of the perfect SDE is as follows. The ground state supports a non-zero current $I^{\text{eq}}$ through
the BAM wire mediated by $q=0$ Cooper pairs in the parent SC. Thus,
driving a different dissipationless current through the BAM wire will
require $q\neq0$ Cooper pairs. Explicitly, the generalizations of
Eqs. (\ref{eq:H-Delta}) and (\ref{eq:current}) are
\begin{align}
H_{k}^{\Delta}(q) & =\left(\begin{array}{cc}
\varepsilon_{k+q} & \Delta_{q}^{*}\\
\Delta_{q} & -\varepsilon_{-k}
\end{array}\right)\\
I(q) & =\intop_{k}\text{Tr}\left\{ j_{k}(q)\left(f\left[H_{k}^{\Delta}(q)\right]-f\left[H_{k}^{0}(q)\right]\right)\right\} 
\end{align}
where $\Delta_{q}$ is the pairing amplitude associated with momentum-$q$
Cooper pairs and $j_{k}(q)=\frac{e}{2}\left(\begin{array}{cc}
v_{k+q} & 0\\
0 & -v_{-k}
\end{array}\right)\otimes\mathbb{I}_{g}$. Since the parent SC is conventional, we expect $|\Delta_{q}|=|\Delta_{-q}|$
and $\Delta_{q}=0$ when $|q|$ exceeds a critical value $q_{c}$.
For weak pairing, the generalization of Eq. (\ref{eq:current answer})
is
\begin{align}
I(q) & =-ge|\Delta_{q}|^{2}\frac{d}{dq}\intop_{k}\frac{\tanh\left(\frac{\varepsilon_{k+q}}{2T}\right)+\tanh\left(\frac{\varepsilon_{-k}}{2T}\right)}{2\left(\varepsilon_{k+q}+\varepsilon_{-k}\right)}\nonumber \\
 & \equiv-ge|\Delta_{q}|^{2}F'(q)\label{eq:F(q)}
\end{align}
Note that $F(q)$ depends purely on the normal state band structure.
It peaks when $q$ is such that $k$ and $k+q$ are distinct Fermi
points. Consequently, $F'(q)$ changes sign when $q$ connects a pair
of Fermi momenta. For instance, in a minimal 1D dispersion with a
single left (right) mover with Fermi momentum $-K_{L}$ ($K_{R}$),
$F'(q^{*})=0$ where $q^{*}=K_{R}-K_{L}$. We demonstrate this property
in Fig. \ref{fig:SDE}(c,d) for a lattice dispersion of the form $\varepsilon_{k}=-2t\cos k-2t'\sin(2k+\theta)-\mu$,
which corresponds to ordinary nearest neighbor and complex second
neighbor hopping. $F(q)$ is analyzed more closely in Appendix \ref{app:Fq}.

The implication of this behavior for the SDE, illustrated in Fig.
\ref{fig:SDE}(e,f), is the following. If $q_{c}<|q^{*}|$, a current
$I$ such that $0<I<I^{\text{eq}}$ (we choose the convention $I^{\text{eq}}>0$)
will be non-dissipative and be carried by Cooper pairs in the BAM
with the appropriate $q$, whereas no value of $q$ can accommodate
a negative supercurrent. Thus, the critical currents are $I_{c}^{+}=I^{\text{eq}}$
and $I_{c}^{-}=0$, and the diode coefficient $\eta=\frac{I_{c}^{+}-I_{c}^{-}}{I_{c}^{+}+I_{c}^{-}}=1$. 

We emphasize that this reasoning for the perfect SDE is immune to the specific form of $\Delta_{q}$ as long it vanishes beyond a critical value of $|q|$. In fact, the perfect SDE will persist
even if the parent SC inherits a slight asymmetry due to the BAM and
acquires unequal critical momenta, $q_{c}^{+}\neq q_{c}^{-}$, provided
$\left|q^{*}\right|>\max q_{c}^{\pm}$. Then, $I(q)$ has the same
sign $\forall q$ for which $\Delta_{q}\neq0$.

On the other hand, if $q_{c}>|q^{*}|$, then Cooper pairs with momentum
$q$ such that $|q^{*}|<|q|<q_{c}$ will enable negative supercurrents
in the BAM and yield a diode coefficient $0<\eta<1$. Naively, if a supercurrent vanishes at a certain $q$ and has a negative slope at that point -- as it happens for $q=q^*$ in Fig. \ref{fig:SDE}(f) -- the superconducting phase is rendered unstable as the the corresponding free energy reaches a local maximum or saddle point \cite{yerin2023multipleq}. However, this is only true in intrinsic SCs; for proximity-induced superconductivity in the BASC, the parent SC effectively provides a training field in Nambu pseudospin space that creates a non-zero $\Delta(q)$ proportional to the pairing amplitude in the parent SC. As a result, the BASC does not extremize the free energy of an isolated BAM and can remain stable as long as the parent SC is well-behaved.

\section{Topological defects}\label{sec:defects}

In $d$ dimensions, a BAM would naturally
be described by a vector order parameter with symmetries of velocity
or momentum. For instance, at low energies compared to the bandwidth,
an intuitive choice for an order parameter is the average Fermi momentum.
Alternately, a real-space quantity with the same symmetries as $\boldsymbol{Q}$
that has gained recent interest is the toroidal moment, $\sim\boldsymbol{r}\times\boldsymbol{m}$,
where $\boldsymbol{r}$ is a position vector and $\boldsymbol{m}$
is a magnetic moment \citep{Ederer2007,Hayami2014,Hayami:2016aa,Hayami_2015,Saito:2018aa,Spaldin2008,Urru:2020aa}.
Yet another choice a spin chiral order along a preferred direction
\citep{Chang2020DSSCO,Hosur_2020}. On purely symmetry grounds, $\boldsymbol{Q}$
can couple to electromagnetic fields as $\boldsymbol{Q}\cdot(\boldsymbol{E}\times\boldsymbol{B})$
and hence, can be trained by mutually perpendicular electric and magnetic
fields. We now investigate the effects of topological defects in $\boldsymbol{Q}$
on the equilibrium current density $\boldsymbol{J}^{\text{eq}}$ and demonstrate an anomalous non-conservation of electric charge. 

\begin{figure}
\subfloat[\label{fig:Domain-wall}]{\includegraphics[width=0.45\columnwidth]{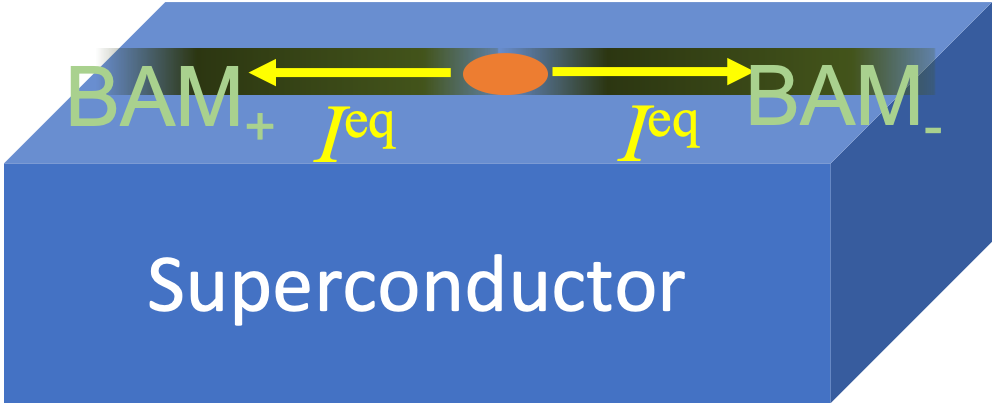}}
\subfloat[\label{fig:Two-domain-walls}]{\includegraphics[width=0.45\columnwidth]{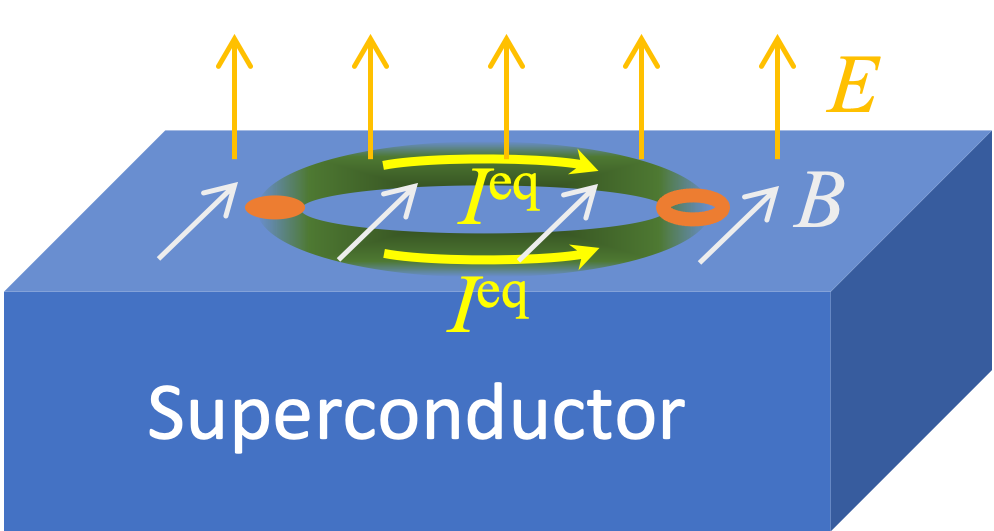}}

\subfloat[\label{fig:E-vortex}]{\includegraphics[width=0.45\columnwidth]{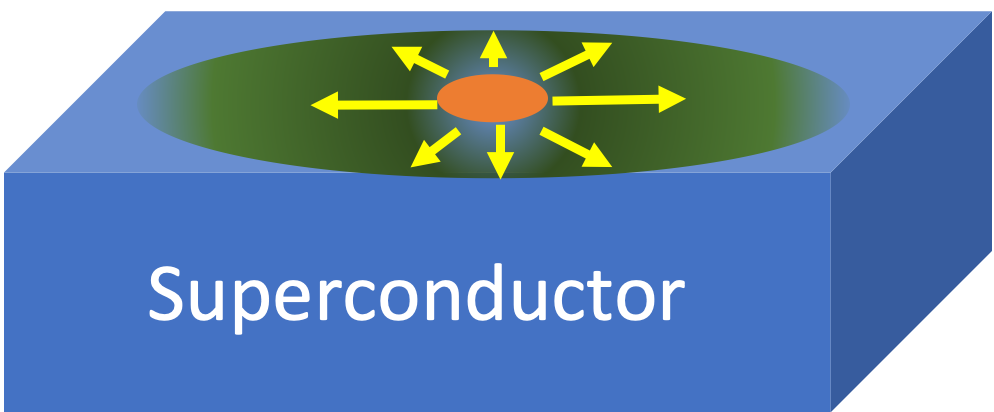}}
\subfloat[\label{fig:M-vortex}]{\includegraphics[width=0.45\columnwidth]{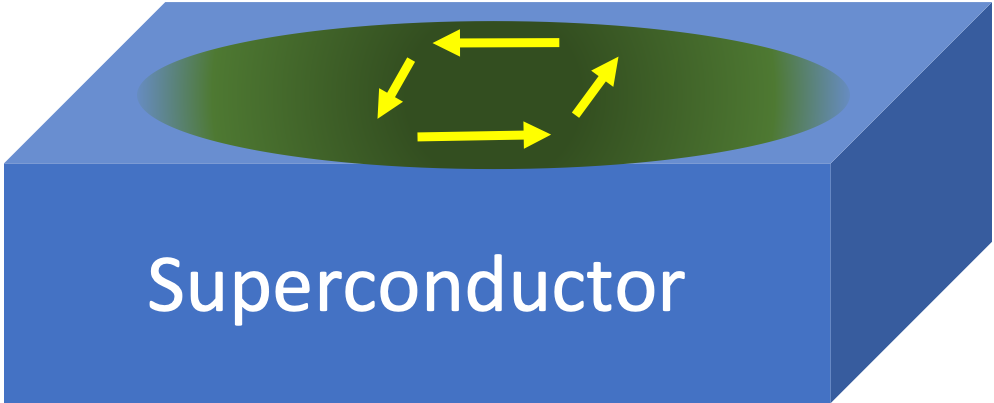}}
\caption{(a) Domain wall (orange dot) in the BAM leads to charge non-conservation
in the BASC due to opposite $I^{\text{eq}}$ on either side of the
domain wall. Here, BAM$_{\pm}$ schematically denote 1D BAMs with
opposite $Q_{x}$. (b) Applying a spatially uniform $\boldsymbol{E}\times\boldsymbol{B}$
on a ring geometry drives currents in the same average direction in
two halves of the ring (left to right in the figure) which results
in a pair of domain walls that emit (orange dot) and absorb (orange
circle) charge. (c) An ``electric'' vortex with $\boldsymbol{\nabla}\cdot\boldsymbol{Q}\protect\neq0$
also acts as a charge source or sink based on the sign of $\boldsymbol{\nabla}\cdot\boldsymbol{Q}$
while (d) a ``magnetic'' vortex with $\boldsymbol{\nabla}\cdot\boldsymbol{Q}=0$
does not exhibit charge non-conservation. \label{fig:Defects}}
\end{figure}

Suppose $\boldsymbol{Q}$ has a domain wall, $\boldsymbol{Q}(\boldsymbol{r})=Q_{0}\Theta(x)\hat{\mathbf{x}}$.
Then, the $x>0$ and $x<0$ regions will carry opposite $I^{\text{eq}}$
(Fig.~\ref{fig:Domain-wall}), so $x=0$ will be a source or sink
of electric charge depending on the directions of $I^{\text{eq}}$
in the two regions. Interestingly, a uniform $\boldsymbol{E}\times\boldsymbol{B}$
field will create a pair of domain walls which will result in a charge
source and a sink in the BASC (Fig.~\ref{fig:Two-domain-walls}).
Similarly, an ``electric vortex'' $\boldsymbol{Q}(\boldsymbol{r})=Q_{0}\frac{x\hat{\mathbf{x}}+y\hat{\mathbf{y}}}{\sqrt{x^{2}+y^{2}}}$
also acts as a source/sink of charge (Fig.~\ref{fig:E-vortex}),
but a ``magnetic vortex'' $\boldsymbol{Q}(\boldsymbol{r})=Q_{0}\frac{y\hat{\mathbf{x}}-x\hat{\mathbf{y}}}{\sqrt{x^{2}+y^{2}}}$
does not (Fig.~\ref{fig:M-vortex}). In general, time-independence
of physical quantities at equilibrium means the continuity equation
for charge conservation is violated when
\begin{equation}
\frac{\partial\rho}{\partial t}+\boldsymbol{\nabla}\cdot\boldsymbol{J}^{\text{eq}}\propto\boldsymbol{\nabla}\cdot\boldsymbol{Q}\neq0
\end{equation}
This is an anomalous charge non-conservation that resembles the chiral
anomaly in 1D quantum Hall edges and 3D Weyl semimetals \citep{Hosur2013a,Hu:2019aa,Burkov2018,Maeda2DAnomaly,NielsenABJ,Wang2017,YanFelserReview,ArmitageWeylDiracReview,Bulmash2015}.
Like the chiral anomaly, charge depleted from one region appears in
a different region that could be macroscopically far away. Also, the
violation is enabled by a ``charge reservoir'' in both cases --
the parent SC for the BASC and a bulk insulator for the chiral anomaly.
On the other hand, the charge non-conservation here differs from that
in the chiral anomaly in crucial ways. Firstly, it can occur in any
number of dimensions in principle -- including 2D, where a chiral
anomaly is absent. Moreover, it depends on material details whereas
the chiral anomaly is determined by universal constants $e$ and $\hbar$. 

Such anomalous non-conservation of charge does not occur in isolated
SCs. Even though their mean-field condensates violate charge conservation,
the nature of the violation is very different. In particular, they
break gauge symmetry spontaneously, which leads to charge non-conserving
microscopic processes such as Andreev reflection at an interface with
a non-superconducting material. However, the actual bulk material
still conserves particle number and obeys the continuity equation.
Inhomogeneities do create local currents in the equilibrium state,
for example, around SC vortices. However, the continuity equation
and time-independence of equilibrium ensure that such currents necessarily
form loops and are divergence-free. Thus, they do not contain local
sources or sinks or charge, let alone a mechanism for pumping charge
non-locally over macroscopic distances. The latter is a unique property
of BASCs.

\section{Experimental platforms}\label{sec:platforms}

BAMs are the generic low energy limit
of any metal that breaks $\mathcal{T}$ and $\mathcal{I}$. This encompasses
well-studied systems where $\mathcal{I}$ is broken by spin-orbit
coupling or ferroelectricity and $\mathcal{T}$ is broken by a Zeeman
field or magnetic order. It also includes an emerging family of metals
with toroidal order such as UNi$_{4}$B, where $\mathcal{T}$ and
$\mathcal{I}$ are broken by a vector order parameter but $\mathcal{TI}$
is preserved \citep{Ederer2007,Hayami2014,Hayami:2016aa,Hayami_2015,Saito:2018aa,Spaldin2008,Urru:2020aa}.
Finally, itinerant electrons in the background of certain unidirectional
spin chiral orders are BAMs too \citep{Hosur_2020,Chang2020DSSCO}.
In principle, any of these systems proximity-coupled to a conventional
SC should exhibit the phenomena discussed in this paper, as the essential
ingredient is an order parameter with the symmetries of $\boldsymbol{E}\times\boldsymbol{B}$.
However, determining the ideal platform with a large effect will require
a more sophisticated study of the proximity effect from a parent SC
that is left for future work.

\section{Summary}\label{sec:summary}

We have studied proximity-induced conventional superconductivity
in metals with asymmetric dispersions, which can be viewed as the
low-energy limit of generic metals that break $\mathcal{T}$- and
$\mathcal{I}$-symmetries. We showed that the resulting SC carries
a persistent equilibrium supercurrent that causes topological defects
in the band asymmetry to act as sources and sinks of charge, both
of which are absent in isolated SCs. For large enough band asymmetry,
the SC also exhibits a perfect SDE while smaller asymmetry still gives
a non-zero SDE. This work reveals reveals strange behaviors in systems
that would normally be considered ``ordinary'', as they are at equilibrium,
lack band Berry phases and acquire conventional, $s$-wave superconductivity.
\begin{acknowledgments}
We acknowledge useful discussions with Lei Hao, John Wei, Kai Chen
and Bishnu Karki. This work was supported by the Department of Energy
under grant no. DE-SC0022264.
\end{acknowledgments}

\bibliographystyle{apsrev4-2}
\bibliography{library}

\appendix
\begin{widetext}

\section{Derivation of $I^{\text{eq}}$ and $I(q)$} \label{app:Iq}

\subsection{$I^{\text{eq}}$}

We begin with 
\begin{equation}
I^{\text{eq}}=\intop_{k}\text{Tr}\left\{ j_{k}\left[f\left(H_{k}^{\Delta}\right)-f\left(H_{k}^{0}\right)\right]\right\} 
\end{equation}
where $H_{k}^{\Delta}=\left(\begin{array}{cc}
\varepsilon_{k} & \Delta_{0}^{*}\\
\Delta_{0} & -\varepsilon_{-k}
\end{array}\right)$, $j_{k}=\frac{e}{2}\left(\begin{array}{cc}
v_{k} & 0\\
0 & -v_{-k}
\end{array}\right)\otimes\mathbb{I}_{g}$ is the current operator, $\intop_{k}\equiv\int\frac{dk}{2\pi}$,
$f(X)=\left[e^{X/T}+1\right]^{-1}$. It is easiest to evaluate the
trace in the eigenbasis of $H_{k}^{\Delta}$. This gives
\begin{align}
I^{\text{eq}} & =\frac{ge}{4}\intop_{k}\sum_{n=\pm}f(E_{k}^{n})\left\langle \psi_{k}^{n}\left|(v_{k}-v_{-k})+(v_{k}+v_{-k})\tau_{z}\right|\psi_{k}^{n}\right\rangle -\left(\Delta=0\text{ contribution}\right)\\
 & =\frac{ge}{4}\intop_{k}\sum_{n=\pm}f(E_{k}^{n})\left[(v_{k}-v_{-k})+\frac{n(v_{k}+v_{-k})}{\sqrt{1+\left|\frac{2\Delta_{0}}{\varepsilon_{k}+\varepsilon_{-k}}\right|^{2}}}\right]-\left(\Delta=0\text{ contribution}\right)
\end{align}
where $\tau_{z}$ is a Nambu Pauli matrix and and $H_{k}^{\Delta}|\psi_{k}^{\pm}\rangle=E_{k}^{\pm}|\psi_{k}^{\pm}\rangle$
with $E_{k}^{\pm}=\frac{\varepsilon_{k}-\varepsilon_{-k}}{2}\pm\text{sgn}\left(\varepsilon_{k}+\varepsilon_{-k}\right)\sqrt{\left(\frac{\varepsilon_{k}+\varepsilon_{-k}}{2}\right)^{2}+|\Delta_{0}|^{2}}$.
Thanks to particle-hole symmetry, $E_{k}^{\pm}=-E_{-k}^{\mp}$, changing
$k\to-k$ in the $n=-$ term simplifies this to
\begin{equation}
I^{\text{eq}}=-\frac{ge}{4}\intop_{k}\tanh\left(\frac{E_{k}^{+}}{2T}\right)\left[(v_{k}-v_{-k})+\frac{v_{k}+v_{-k}}{\sqrt{1+\left|\frac{2\Delta_{0}}{\varepsilon_{k}+\varepsilon_{-k}}\right|^{2}}}\right]-\left(\Delta=0\text{ contribution}\right)
\end{equation}
To leading order in $\Delta_{0}$, 
\begin{equation}
I^{\text{eq}}=-\frac{ge|\Delta_{0}|^{2}}{2}\intop_{k}\left[\frac{v_{k}}{2T(\varepsilon_{k}+\varepsilon_{-k})}\text{sech}^{2}\left(\frac{\varepsilon_{k}}{2T}\right)-\frac{v_{k}+v_{-k}}{(\varepsilon_{k}+\varepsilon_{-k})^{2}}\tanh\left(\frac{\varepsilon_{k}}{2T}\right)\right]
\end{equation}
Integrating the first term by parts further reduces this to the expression in the main paper:
\begin{equation}
I^{\text{eq}}=ge|\Delta_{0}|^{2}\intop_{k}\frac{v_{-k}}{(\varepsilon_{k}+\varepsilon_{-k})^{2}}\tanh\left(\frac{\varepsilon_{k}}{2T}\right)
\end{equation}

\subsection{$I(q)$}

The starting point now is
\begin{align}
H_{k}^{\Delta}(q) & =\left(\begin{array}{cc}
\varepsilon_{k+q} & \Delta_{q}^{*}\\
\Delta_{q} & -\varepsilon_{-k}
\end{array}\right)\\
I(q) & =\intop_{k}\text{Tr}\left\{ j_{k}(q)\left(f\left[H_{k}^{\Delta}(q)\right]-f\left[H_{k}^{0}(q)\right]\right)\right\} 
\end{align}
where $j_{k}(q)=\frac{e}{2}\left(\begin{array}{cc}
v_{k+q} & 0\\
0 & -v_{-k}
\end{array}\right)\otimes\mathbb{I}_{g}$. Particle-hole symmetry now reads $E_{k+q/2}^{\pm}=-E_{-k+q/2}^{\mp}$.
Thus, we shift $k\to k-q/2$ and change $k\to-k$
in the $n=-$ term. Paralleling the steps used for $I^{\text{eq}}$
yields,
\begin{equation}
I(q)=ge\left|\Delta_{q}\right|^{2}\intop_{k}\frac{v_{-k+q/2}\tanh\left(\frac{\varepsilon_{k+q/2}}{2T}\right)}{(\varepsilon_{k+q/2}+\varepsilon_{-k+q/2})^{2}}
\end{equation}
to leading order in $\Delta_{q}$, which clearly equals $I^{\text{eq}}$
at $q=0$. To shed light on the $q$-dependence, we use the freedom
in redefining the integration variable $k$ to rewrite $I(q)$ as
\begin{align}
I(q) & =ge\left|\Delta_{q}\right|^{2}\intop_{k}\frac{v_{-k+q}\tanh\left(\frac{\varepsilon_{k}}{2T}\right)}{(\varepsilon_{k}+\varepsilon_{-k+q})^{2}}\\
 & =-ge\left|\Delta_{q}\right|^{2}\frac{d}{dq}\intop_{k}\frac{\tanh\left(\frac{\varepsilon_{k}}{2T}\right)}{\varepsilon_{k}+\varepsilon_{-k+q}}\\
 & =-\frac{ge\left|\Delta_{q}\right|^{2}}{2}\frac{d}{dq}\intop_{k}\frac{\tanh\left(\frac{\varepsilon_{k+q}}{2T}\right)+\tanh\left(\frac{\varepsilon_{-k}}{2T}\right)}{\varepsilon_{k+q}+\varepsilon_{-k}}\\
 & =-ge\left|\Delta_{q}\right|^{2}F'(q)
\end{align}
as given in the main paper.

\section{Evaluation of $F(q)$}\label{app:Fq}

We have
\begin{align}
F(q) & =\frac{1}{4T}\intop_{k}\frac{\tanh\left(\frac{\varepsilon_{k+q/2}}{2T}\right)+\tanh\left(\frac{\varepsilon_{-k+q/2}}{2T}\right)}{\frac{\varepsilon_{k+q/2}+\varepsilon_{-k+q/2}}{2T}}\\
 & =T\intop_{k}\frac{\sinh\left(\frac{\varepsilon_{k+q/2}+\varepsilon_{-k+q/2}}{2T}\right)}{(\varepsilon_{k+q/2}+\varepsilon_{-k+q/2})\left[\cosh\left(\frac{\varepsilon_{k+q/2}+\varepsilon_{-k+q/2}}{2T}\right)+\cosh\left(\frac{\varepsilon_{k+q/2}-\varepsilon_{-k+q/2}}{2T}\right)\right]}
\end{align}
This integral is sharply peaked at $\varepsilon_{k+q/2}=\varepsilon_{-k+q/2}$.
Let us refer to the points that satisfy $\varepsilon_{k+q/2}=\varepsilon_{-k+q/2}$
as $K_{i}$ and the corresponding energies as $\varepsilon_{i}$.
Note that $K_{i}$ and $\varepsilon_{i}$ depend on $q$. We can evaluate
$F(q)$ using Laplace's method by defining 
\begin{align}
G_{q}(k) & =\ln\left\{ \frac{\tanh\left(\frac{\varepsilon_{k+q/2}}{2T}\right)+\tanh\left(\frac{\varepsilon_{-k+q/2}}{2T}\right)}{\frac{\varepsilon_{k+q/2}+\varepsilon_{-k+q/2}}{2T}}\right\} 
\end{align}
so that
\begin{equation}
F(q)\approx\sum_{i}\sqrt{\frac{2\pi}{\left|G_{q}^{\prime\prime}(K_{i})\right|}}\frac{\tanh\frac{\varepsilon_{i}}{2T}}{2\varepsilon_{i}}
\end{equation}
Straightforward algebra yields
\begin{align}
G_{q}^{\prime\prime}(K_{i}) & =\frac{v_{K_{i}+q/2}^{\prime}+v_{-K_{i}+q/2}^{\prime}}{2T}\left[\frac{1}{\sinh\left(\frac{\varepsilon_{i}}{T}\right)}-\frac{1}{\left(\frac{\varepsilon_{i}}{T}\right)}\right]\\
 & -\frac{v_{K_{i}+q/2}^{2}+v_{-K_{i}+q/2}^{2}}{4T^{2}\cosh^{2}\left(\frac{\varepsilon_{i}}{2T}\right)} +\left(\frac{v_{K_{i}+q/2}-v_{-K_{i}+q/2}}{2T}\right)^{2}\left[\frac{1}{\left(\frac{\varepsilon_{i}}{T}\right)^{2}}-\frac{1}{\sinh^{2}\left(\frac{\varepsilon_{i}}{T}\right)}\right]\nonumber\\
 & \approx-\frac{v_{K_{i}+q/2}^{\prime}+v_{-K_{i}+q/2}^{\prime}}{2\varepsilon_{i}}\text{ for }|\varepsilon_{i}|\gg T,\left|q(v_{K_{i}}-v_{-K_{i}})\right|
\end{align}
Thus,
\begin{equation}
F(q)\approx\sum_{i}\sqrt{\frac{\pi}{\left|\varepsilon_{i}\left(v_{K_{i}+q/2}^{\prime}+v_{-K_{i}+q/2}^{\prime}\right)\right|}}\text{ for }|\varepsilon_{i}|\gg T,\left|q(v_{K_{i}}-v_{-K_{i}})\right|
\end{equation}
Clearly, $F(q)$ is maximum when $\varepsilon_{i}$, i.e., the momenta
$K_{i}$ are Fermi momenta. At larger $\varepsilon_{i}$, it decays
as $1/\sqrt{\varepsilon_{i}}$. 

\end{widetext}
\end{document}